\documentclass[12pt,a4paper]{revtex4}

\usepackage{epsfig}
\newcommand{\beq}{\begin{eqnarray}}
\newcommand{\eeq}{\end{eqnarray}}  

\begin{document}
\textheight= 8.0 in
\topmargin= 0.25 in
\rightmargin= 1 in
\leftmargin= 1 in
\baselineskip=10 pt
\baselineskip=2\baselineskip
\vskip 3cm

\title{Modeling, Performance Analysis and Comparison of Two Level Single 
Chain Pointer Forwarding Strategy For Location Management in Wireless Mobile 
Communication.}

\author{Chhaya Ravi Kant}
 \affiliation{Department of Physics, Indira Gandhi Institute of Technology, 
Guru Gobind Singh Indraprastha University, Delhi 110 006, India.}
  \email{chhaya_rkant@yahoo.co.in}
\author{P. Arun}
 \affiliation{Department of Physics \& Electronics, S.G.T.B. Khalsa College,
University of Delhi, Delhi - 110 007, India}
  \email{arunp92@physics.du.ac.in, arunp92@yahoo.co.in }
\author{Nupur Prakash}
 \affiliation{School of Information technology, Guru Gobind Singh
Indraprastha University, Delhi 110 006, India.}
  \email{nupurprakash@rediffmail.com}

\maketitle
\vskip -0.5cm
\noindent \rule[0.1in]{6.5in}{0.5mm}
\vskip 0.25cm
\section*{Abstract}

Global wireless networks enable mobile users to communicate regardless of their
locations. Location management is an important part of the emerging wireless
and mobile technology. A Personal Communication System (PCS) network must have 
an efficient way to keep track of the mobile users to deliver services 
effectively. Global System for Mobile Communication (GSM) is a commonly 
accepted standard for mobility management of mobile users. Location management 
involves location tracking, and location information storage. Location
management requires mobile users to register at various registration 
areas whenever they are on the move. The registration process may cause excessive 
signaling traffic and long service delays. To improve the efficiency of location 
tracking and avoid call set up delays, several strategies such as local anchor  
scheme, per-user caching scheme and several pointer forwarding schemes have been 
proposed in the past. In this paper, we propose a new {\bf Two Level Single 
Chain Pointer Forwarding (TLSCP) Strategy} in which a two level hierarchy of 
level-1 
and level-2 forwarding pointers reduced to single chain length are used. 
Organizing the pointers in a two level hierarchy and further restricting the 
pointer chain lengths at both the levels to single chain localizes the network 
signaling traffic and reduces the call set up delays.
To justify the effectiveness of our proposed strategy, we develop an analytical
model to evaluate the signaling cost. Our performance analysis shows that the
proposed dynamic TLSCP scheme can significantly reduce the network signaling 
traffic and cost for different categories of users under different network 
conditions 


\section{Introduction}

The phenomenal growth in cellular telephony over the past several years has
demonstrated the vitality of mobile communication for users on the
move \cite{c1}. 
Infact, in a decade of it's existence, the number of mobile phone owners in
Delhi is greater than the number of people who own land line phones. If this
is true for a developing country, it maybe safe to assume the same for urban
areas of developed countries. One reason for mobile usage gaining popularity 
is the rapid fall in it's cost. Along with the competition in the electronics 
industry driving the hardware cost down, improvement in strategies to locate 
a mobile user and transferring the calls have led to reduction in operational 
cost. Location management is an important part of the emerging wireless mobile 
communication system \cite{c2}. A network must retain information about location 
of users in the network in order to route traffic to the correct destination. 
Location Management has two components: 
\begin{itemize}
\item{} Location tracking, and 
\item{} Location information storage. 
\end{itemize}
Location tracking mechanisms may be perceived as updating and querying a 
location database of users to determine when and how a change in location 
database entry should be initiated. Location information storage mechanisms 
help in organizing and maintaining the location database. Location tracking 
typically consists of two operations:
\begin{itemize}
\item{(i)} updating (or  Registration), the process by which a mobile user
initiates a change in the location database according to its new location,
and
\item{(ii)} Finding (or Paging), the process by which the network initiates a 
query for a mobile user location (which may also result in an update to the 
location database). 
\end{itemize}
Two popular and very similar schemes for mobility management 
are EIA/TIA Interim Standard IS-41 \cite{c3} and GSM Mobile application Part 
(MAP) \cite{c4,c5}. However both these schemes use a two tier system of a Home 
Location 
Register (HLR) and a Visitor Location Register (VLR) databases and share the 
same location update procedure. The service area is divided into several 
registration areas. When a user subscribes to the service, a record or a 
profile of the user is kept in the HLR, a register located close to the Mobile 
Switching Center (MSC). When the mobile host moves to a new location or out of 
the coverage of this MSC, it informs the VLR in the new area and a temporary 
record is created in the VLR. The VLR sends a registration message to the HLR, 
which in turn responds with necessary user information. When a call arrives for 
the mobile user, HLR is queried which in turn submits a query to the current VLR 
for the current rout-able address of the mobile host. According to these 
strategies \cite{c4,c5}, a mobile user performs location update (registration) 
at the HLR every time the user crosses the boundary of a Registration area (RA) 
and De-registers at the previous VLR. Thus, the registration process incurs high 
signaling traffic when many users cross their RAs. This problem worsens with 
increase in the number of users and when many users are far away from their HLR. 
To overcome this problem, location tracking techniques should use a combination 
of updating and finding in an effort to select the best trade-off between update 
overhead and delay incurred in finding the host. A trade-off also needs to be 
analyzed between the update and paging costs \cite{c6}. The network signaling 
traffic is also dependent on the geographical location of the HLR databases as it 
affects both latency and overhead when location information is processed
\cite{c1}. Both of these factors need to be minimized as they affect the network 
performance. Distributed HLRs in several areas can prevent the HLR from 
becoming a bottleneck in signaling network. The latency can be reduced by 
using replicated databases where location information is kept at several places 
in the network. Such user profile replication can reduce HLR access. The amount 
of signaling traffic may increase slightly as a change in location information 
after a move must be initiated in all to modify all replications. Various 
location management strategies for reducing the location management cost have 
been proposed in the last decade \cite{c7}-\cite{c15}. The different proposed 
strategies have been analyzed using different mathematical and analytical models 
\cite{c16}-\cite{c18}. Different mobility models \cite{c16} have been taken 
into account to evaluate the performance of these schemes. In random mobility 
models, a mobile user is likely to move to any one of the neighboring cells, 
where as, in the activity based models the various activities of the users for 
classification are taken into account to decide the updating process. Simple 
Markov models have been utilized to assess and compare the performance 
characteristics of the 'Degradable location management scheme' \cite{c18} using 
the IS-41 protocol, the Forwarding and Resetting algorithm \cite{c11}, the Paging 
and Location Update Algorithm \cite{c16} and the Local Anchor Scheme
\cite{c12}. The proposed 'Local Anchor (LA) scheme' \cite{c12} and 'Per user 
pointer forwarding scheme' \cite{c16} avoid expensive HLR updates and access 
each time a mobile user moves to a new registration area (RA). However, the LA 
scheme \cite{c12} has the drawback of increased local signaling 
traffic whereas the Per-user pointer forwarding \cite{c16} scheme suffers a long 
set-up delay for highly mobile users. An efficient 'Two Level Pointer Forwarding 
(TLPF) Strategy' was proposed by Ma and Fang \cite{c15}, in which a two level 
mobility management was introduced by selecting a set of VLRs as Mobility 
Agents responsible for location management in geographically larger area 
compared to the RA of the VLR. Two kinds of pointers were used between the 
VLRs and MAs. Registration signaling traffic was localized and thus the network 
signaling traffic was considerably reduced. The performance analysis of TLPF 
strategy showed significant reduction in the network signaling traffic for 
users with low Call to mobility (CMR) with slight increase in the call set-up 
delay. To overcome the network signaling and call setup delay problems, we 
proposed a 'Two Level Single Chain Pointer Forwarding (TLSCP) strategy'
\cite{c19, c20}. 

\par Similar to the TLPF strategy of Ma \cite{c15}, in the 
presently proposed TLSCP strategy, a two level hierarchy of level-1 and 
level-2 forwarding pointers has been used. 
However, in the proposed strategy, the pointer chain length of both level-1 and 
level-2 pointers chains has always been adjusted and reduced to single chain. 
Organizing the pointers in a two level hierarchy and further restricting the 
pointer chain lengths at both levels to single chain reduces the network 
signaling traffic and call set-up delays. We have further combined the concept 
of two level single chain hierarchy with choice of replicated HLRs distributed
throughout the geographic region. The use of forwarding pointers makes the 
concept of distributed HLRs more attractive and efficient. This makes the 
proposed-TLSCP Strategy an efficient and attractive scheme for location 
management of mobiles. In the present paper, we have done the modeling and 
performance analysis of the TLSCP scheme using an analytic model. We have also 
made a performance comparison of the proposed scheme with the Basic GSM
\cite{c4, c5} and the TLPF strategy proposed be Ma and Fang \cite{c15}. In the 
next section, we describe the basic architecture of the distribution of VLRs 
and HLRs assumed in the proposed TLSCP strategy to facilitate the presentation 
and analysis of the strategy. Section III gives the pseudo-code of the Basic 
GSM strategy and the proposed TLSCP strategy. In section IV, we present 
an analytic model of the performance of the TLSCP strategy. Comparison of the
performance of TLSCP strategy has been made with the performance of TLPF 
strategy \cite{c15} as well as the basic GSM scheme and the results have been 
summarized in section V. Finally in Section VI we present the conclusion and 
the future scope of our work. 

\section{Two Level Single Chain Location Pointer (TLSCP) Management Strategy}

The proposed TLSCP scheme is based on PCS architecture. The whole geographic
area is divided into a number of sub areas. There are a number of registration 
areas (RAs) in each sub-area. Each RA has a VLR database maintaining the records 
of all mobile users present in the RA. We have proposed to dedicate one 
centrally located VLR in each sub-area to act as Mobility Agent (MA), 
responsible for location management of all the mobile users present in 
the RAs lying in the coverage area of that MA. A hexagonal network coverage 
model \cite{c12} for VLRs has been assumed in which Visitor Location Registers 
(VLRs) are arranged in rings around the centrally located MA. Thus each MA 
covers an n-layer VLR structure and would be responsible for location 
management in a large area. We have proposed a 
replicated and distributed HLR environment throughout the geographic region. 
Each distributed HLR maintains one record for an authorized mobile Each HLR is 
located at the 
site of Public Switching Telephone Network (PSTN). This offers an advantage 
that when an incoming call is originated from some remote PSTN, the HLR would 
be queried directly and Global Title Translation (GTT) would not be required 
because the distributed HLR is near the Signaling Transfer Point (STP)
\cite{c21}.

\begin{figure}[htb]
\epsfig{file=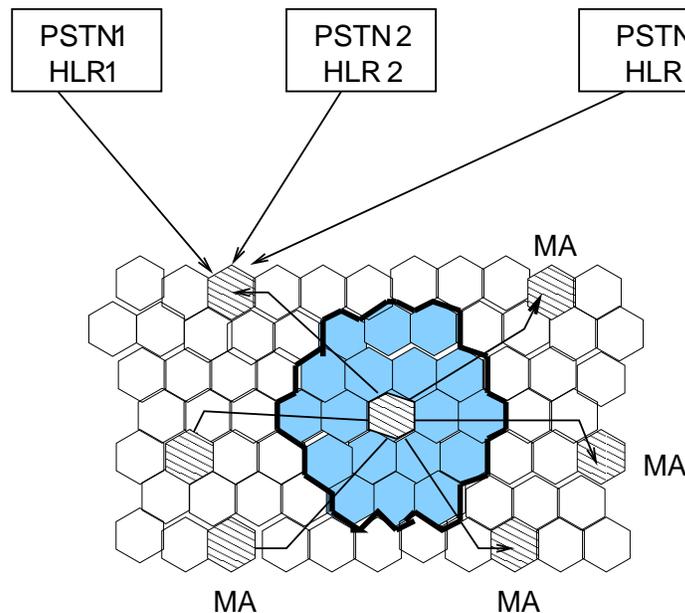, width=4in}
\vskip -0.5cm
\caption{\sl A geographical region is divided into numerous hexagonal area
called registration area (RA). Various RAs are bought together (example
shaded 19 hexas) and a centrally located VLR is assigned the role of
mobility agent (MA).}
\end{figure}

As mentioned earlier, each distributed HLR has one record for an authorized
mobile user. This record contains the individual MA-id where the mobile was 
found last time by an individual HLR. Different HLRs may record different MAs/ 
sub-areas as the head of their own locating paths. So, each HLR may have a 
different locating path for the same mobile. Thus one cellular system with x 
distributed HLRs may have y different locating pointer chains (where y$\leq$x) 
for the same mobile user. Further, as a mobile user moves from one RA to another, 
a level-2 pointer chain is set up between VLRs of different RAs and further a 
level-1 pointer chain is set up between MAs if the new RA lies in the coverage 
area of another MA. The lengths of both level-1 and level-2 pointer chains are
adjusted and reduced to single length between MAs and VLRs respectively. 
This constraint is always true regardless of the HLR from which an incoming 
call originates. Based on this characteristic, the proposed scheme is called as,
'Two level single chain pointer forwarding strategy' for location tracking in a 
distributed HLR environment.

\section{OPERATIONS DEFINED IN TLSCP SCHEME}

The two basic procedures in location management of mobile users are
\begin{itemize}
\item{Move( )-} Defines movement of mobile user from one Registration Area (RA) 
to another. 
\item{Locate( )-} Determines the location of RA where the user is currently 
located.
\end{itemize} 
The pseudo-code for the two basic operations defined for the GSM and the
proposed TLSCP strategy are as follows \cite{c19}\\
GSM MOVE( )\\
\{\\
The mobile terminal detects that it is in a new registration area;\\
The mobile terminal sends a registration message to the user's HLR;\\
The HLR sends a registration cancellation message to the old VLR;\\  
The old VLR sends a cancellation confirmation message to the HLR;\\  
The HLR sends a registration confirmation message to the new VLR;\\
\}\\   
GSM LOCATE( )\\
\{\\
Call to a PCS user is detected at the local switch;\\
If the called party is in the same RA, then return;\\
Switch queries the called party's HLR;\\
HLR queries the called party's current VLR, V;\\
VLR, V returns the called party's location to HLR;\\
HLR returns the location to the calling party;\\
\}\\
\par The pseudo-codes for the TLSCP Move() and Locate() procedures are given 
below\\
TLSCP-MOVE(  )\\
\{\\ 
The user on moving to the new RA registers at the new VLR.\\
IF NEW-VLR and OLD-VLR both belong to same MA.\\
\{\\     
NEW-VLR deregisters the user OLD-VLR.\\
OLD- VLR sends Acknowledgement-signal to NEW-VLR.\\
OLD-VLR sets up a level-2- pointer to NEW-VLR.\\
OLD-VLR forwards the De-registration signal to other VLRs in level-2 pointer 
chain.\\
Length of level-2 pointer Chain is adjusted to single-chain between two
VLRs.\\
\}\\ 
ELSE\\
\{\\
User registers at NEW-MA.\\
NEW-MA sends De-registration message to OLD-MA. \\
OLD-MA sends Acknowledgement signal to NEW-MA.\\ 
OLD-MA sets up a level-1 pointer to NEW-MA.\\
IF OLD-MA was already a node in level-1 pointer chain.\\
\{\\
OLD-MA forwards the MA- De-registration message to MAs in level-1 pointer
chain;\\ 
Level-1 pointer chain length is adjusted to single length between two MAs;\\
\}\}\}\\
TLSCP-LOCATE ( )\\
\{\\
A call to a user is detected at a PSTN switch; \\
PSTN queries the user's HLR.\\
HLR queries the MA at the head of its locating chain.\\
MA queries the VLR  pointed by the pointer VLR-PTR in its data entry;\\
IF VLR-PTR=NULL\\
\{\\
MA forwards call-request to Next-MA in level-1 MA pointer chain;\\
\}\\
MA forwards call request to the VLR pointed by VLR-PTR.\\
VLR locates the user tracking through the level-2 pointer chain.\\
User's current-VLR sends a rout-able address of user to the MA.\\
MA forwards the rout-able address of user to HLR and PSTN switch.\\
PSTN transfers the call to the user.\\
HLR updates the user's location in its database to current-LN.\\
\}\\  
\par The detailed protocol of the proposed TLSCP strategy as represented by the
pseudo code guarantees that not more than one of each level-1 and level-2 
forwarding pointers will be traced to locate a mobile user regardless of the 
remote PSTN where an incoming call originates. The users have been classified 
into different classes according to their Call-to-Mobility Ratio (CMR). The CMR 
of a user is defined as the expected number of calls for a user during the time 
the user visits an RA, where CMR is defined here in terms of the calls received 
by the user and not the calls originating from the user. Let 
\begin{itemize}
\item{} $\lambda$ be the mean rate at which calls are received by a user.
\item{} 1/$\mu$ be the mean time for which a user resides in a given RA.
\end{itemize}
Then, CMR represented by 'p'
\begin{eqnarray}
p = {\lambda \over \mu}
\end{eqnarray}
Assume that the user crosses several RAs between two consecutive calls. If
the Basic Strategy is used, the HLR is updated every time the user moves to a 
new RA. On the other hand if 'Two level single chain pointer forwarding
strategy' proposed in the paper is used, the HLR is updated only at the rate at 
which a user receives calls. So, HLR is updated $\lambda$ times irrespective of 
the number of moves made by the user. For all the moves if the calls are not 
received only level-1 and level-2 pointer chains are set up and their 
lengths are adjusted. Consider the case in basic mobile telephony, as a mobile 
user moves he registers himself in the 
RA he is present in. The RA updates the MA (several RAs are clustered and
are ascribed under one MA, shaded region of fig 1.) about the residency of the 
user. For routing a 
incoming call to the mobile, the mobile is traced by a pointer pointing between 
the MA and the RA. Consider fig 2, if the user is in RA labeled ${\rm V_1}$, a
pointer would exist between MA and ${\rm V_1}$. A movement into neighboring
RA would result in a new pointer being created between MA and say ${\rm
V_2}$. This creation of pointer is accompanied by various signal transfers
between the MA and RA, creating large signal traffic at MA. To over come
this problem Ho et al \cite{c12} proposed that a RA acts as a Local Anchor
(LA) and any movement within a MA would only result in updating of the LA,
thus cutting down the signal traffic to the MA. However, this is at the
expense of longer pointers to locate an user. For example if a mobile user
goes from ${\rm V_1}$ (with ${\rm V_1}$ acting as the LA) to ${\rm V_2}$ then
${\rm V_3}$ (see fig 2), three pointers would exist, namely MA $\rightarrow$
${\rm V_{1-2}}$ $\rightarrow$ ${\rm V_{2-3}}$. While this cuts down the
traffic of signals to the MA, in locating a mobile user to connect his call,
it would be necessary to traverse the long path MA $\rightarrow$
${\rm V_{1-2}}$ $\rightarrow$ ${\rm V_{2-3}}$. In view of the models
advantages Ma {\it et al} \cite{c15} improved upon the model to cut down
cost. They achieved their objective by changing the local anchor (LA) after
the pointer reached a critical length. The scheme was called "Two Level
Pointer Forwarding Strategy ({\bf TLPFS})".

\begin{figure}[htb]
\epsfig{file=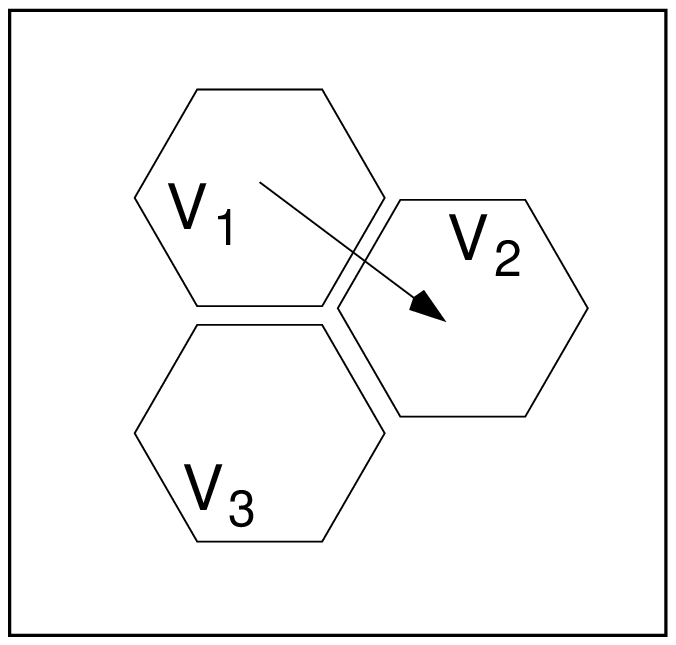, width=2.5in}
\hfil
\epsfig{file=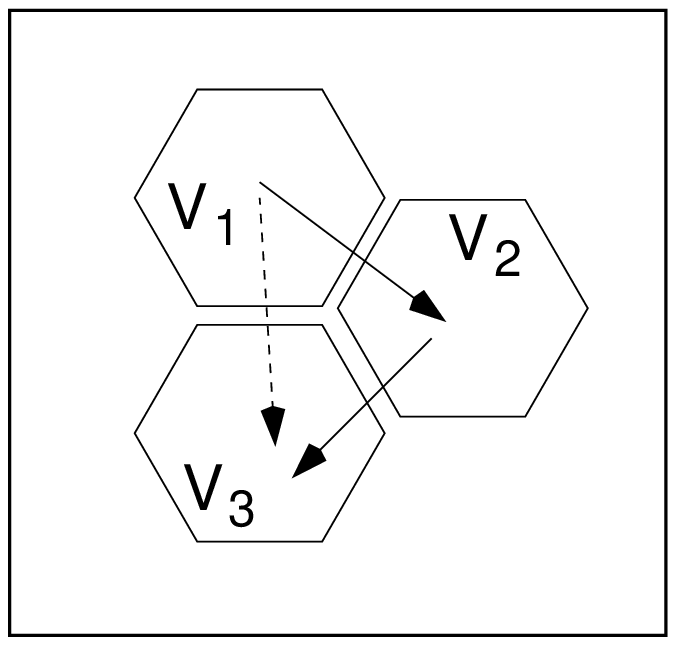, width=2.5in}
\vskip -0.5cm
\caption{\sl Level-2 pointer formation within a MA after one and two steps, 
respectively.}
\end{figure}

\par In this present model (TLSCP), we propose the formation of a resultant 
({\it as in vector}) pointer with every step to a neighboring RA. This reduces 
the length of the pointer as also number of pointer to be traversed while
locating a mobile to transfer a call (now only two pointers have to be
traced). This is at an obvious additional cost of creation of resultant
pointers. For example consider the case shown in fig 2. After the first
step, the pointer for locating a user would be MA $\rightarrow$ ${\rm
V_{1-2}}$ which after the second step would be updated as MA $\rightarrow$ ${\rm
V_{1-3}}$.

\par The following section makes the performance analysis of this
(TLSCP) strategy. The utility and in-turn the success of this TLSCP strategy can
only be estimated by such an analysis.

\section{Performance}

In the following section, we analyze the performance of TLSCP strategy for
mobile technology along the lines of Ma and Fang's works \cite{c15}. At the
end, we compare the performance of this strategy with their Two-level
pointer forwarding strategy. For this we adopt the same notations of their
work with the notations having the same meanings. Adopting these notations
as standard would help in performance analysis of all possible proposed
strategies. However, for brevity and easy reference we detail the notations
and their respective meanings here,\\
\begin{center}
\parbox{5.5in}{
{\sl 
${\rm \alpha (i)}$ = the probability that there are 'i' VLR 
crossings between two consecutive calls.\\
${\rm S_1}$ = the cost of setting up a level 1 pointer between 
two neighboring MAs.\\
${\rm S_2}$ = the cost of setting up a level 2 pointer between 
two neighboring VLAs.\\
}
}
\end{center}

A MA is formed by 'n' closely packed hexagonal RA's. As a mobile user
moving from one RA to another, in a single step, there is a definite
possibility of the user crossing over from one MA to another. While
cross-over from one MA to another MA leads to a creation of a new "{\it
level-1 pointer}", cases where such cross-over does not take place, the
mobile user is only roaming in one of the various RA's of a given MA. This 
activity initiates formation of "{\it level-2 pointers}". The probability of 
remaining in a given MA after one step movement is given as
\beq
P_2={3n^2-5n+2 \over 3n^2-3n+1}\label{e1}
\eeq
Thus, the probability of cross-over, and in-turn formation of level-1
pointer, is given as
\beq
P_1 &=& 1- P_2 = 1-{3n^2-5n+2 \over 3n^2-3n+1}\nonumber\\
&=& {2n-1 \over 3n^2-3n+1}\label{e2}
\eeq
If the mobile user takes 'i' steps, the 
number of steps taken within an MA (j) and the number of steps leading to
crossing of MA boundaries (k) are
\beq
j=i \times P_2=i\left({3n^2-5n+2 \over 3n^2-3n+1}\right)\label{e3}
\eeq
and 
\beq
k=i \times P_1=i\left({2n-1 \over 3n^2-3n+1}\right)\label{e4}
\eeq
respectively. With the knowledge of the probabilities and the total steps
taken between two calls, one only needs an idea of the path the user
undertakes to calculate the cost due to moving from one RA to another and
position movement between MAs. Though such paths would purely be random, one
can specify all possible paths and generalize for an estimation of cost.
This is what we propose to do in the following section.

\subsection{Cost due to Move}
\subsubsection{Cost due to movement within MA}
Each ${Basic\_Move}$ costs the service provider. The general expression for the 
cost involved in level-2 pointer formation covers the new VLR visited by the 
user communicating to the old VLR for informing the users new position. This 
action is also accompanied by De-registration signal and acknowledgment signal 
between all the VLRs the user has occupied since the last call. Thus, with each 
step cost is incurred due to the associated De-registration and acknowledgement. 
The cost incurred would increase with increasing length of the resultant.
Assume a single step movement to the neighboring VLR results in formation
of a pointer (vector) whose length is ${\rm R_1}$ and represents the cost
of pointer formation. For the first step, this would be the resultant
itself. To estimate the cost incurred as a user moves, we assigned the cost
as ${\rm S_2}$ (i.e. the cost incurred on setting pointer ${\rm R_1}$ is
${\rm S_2}$). When the mobile user takes a second step (see fig 2b) another
pointer ${\rm R_1}$ is set representing the step taken to neighboring VLR
(${\rm V_2}$). As per our scheme, a sequence of signalling results giving
rise to single chain pointer ${\rm R_m}$. Again the cost incurred in setting
up this pointer would be proportional to it's length. The length of the
resultant pointer formed would be given as 
\beq
R_m=\sqrt{R_{m-1}^2+S_2^2-2S_2R_{m-1}cos\theta}\label{ee5}
\eeq
where ${\rm \theta}$ is the angle enclosed between the pointer of the latest
step (${\rm R_1}$) and the previous resultant. In the example of fig 2b, the
second step resultant is (${\rm \theta=60^o}$)
\beq
R_2=\sqrt{R_1^2+R_1^2-R_1R_1}=R_1\nonumber
\eeq
The assumption of hexagonal areas defining a VLR reduces the possibilities of 
${\rm \theta}$ to ${\rm 0^o}$ (user steps backward), ${\rm 60^o}$ (user
moves in circles), ${\rm 120^o}$ and ${\rm 180^o}$ (moves radially outward).
One can generalize the cost involved on taking 'j' steps within a MA by
counting all the pointers set up with each step as
\beq
=(j-1)R_1+\sum_{m=1}^{j}R_m\label{e5}
\eeq
\begin{figure}[htb]
\epsfig{file=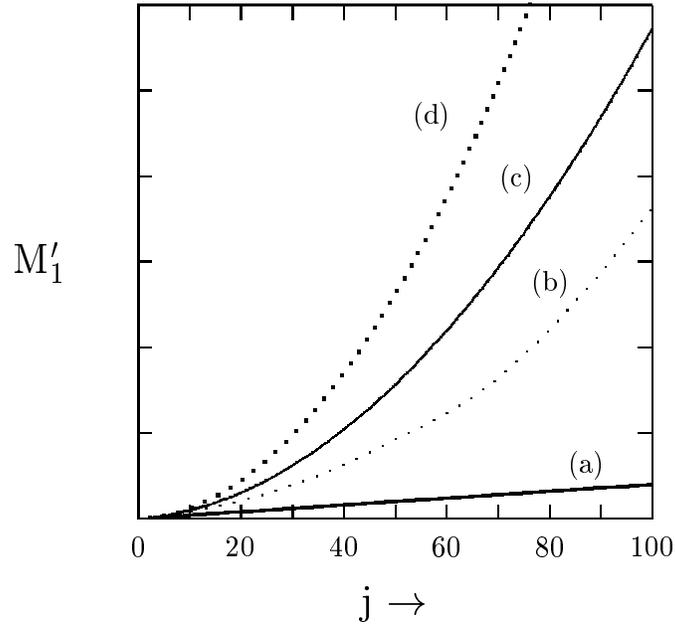, width=3.5in}
\vskip -0.25cm
\caption{\sl Variation in cost incurred on level 2 pointer with increasing steps
taken. The graph is plotted for unity cost, i.e. ${\rm S_2=1}$. Curve (a) is
when second step is taken at ${\rm 60^o}$ w.r.t. previous resultant while
(b) shows cost variation for steps taken at random angles w.r.t. previous
resultants.}
\end{figure}
Figure 3 shows the cost incurred for various paths taken by the user for
increasing number of steps. The curve showing the cost for circular and
radially outward motion encloses all possible paths. Curve (b) shows cost
for a randomized motion (user takes random ${\rm \theta}$ with each step.
This curve was numerically stimulated). As expected this curve lies between
the limiting cases of circular and radially outward motion. Using
eqn(\ref{ee5}) and eqn(\ref{e5}) it can be seen that the cost involved in
setting up level-2 pointer in case a user is moving in circle around the
originating VLR is related to the number of steps taken (j) and is given as
\begin{eqnarray}
=(2j-1)R_1\nonumber
\end{eqnarray}
A similar neat expression is obtained for the user moving radially outward
and is given as
\begin{eqnarray}
=(0.5j^2+1.5j-1)R_1\nonumber
\end{eqnarray}
Since cost incurred by a random motion would lie between curves represented
by these two curves, one can expect a quadratic expression for estimating the 
cost incurred in pointer formation (level-2) due to any random motion after 'j' 
steps as
\beq
&=& (aj^2+bj-1)R_1\nonumber\\
&=& (ai^2P_2^2+biP_2-1)S_2\label{e6a}
\eeq
where simple algebra shows that since curve (b) is constraint to lie between
curve (a) and (d) the coefficient 'a' can take only values between zero and
1/2 and the coefficient 'b' should have values between 1.5 and 2 besides
satisfying the condition a+b=2.
A polynomial of higher order was not selected for simplifying calculations
as also to retain some generality of expression.

\subsubsection{Cost due to movement outside MA}
As the user moves out of a MA, signals of registration and De-registration
are sent between the new VLR and the previously occupied VLR. Recognizing
the new VLR to be existing outside it, the original MA sends information to
the original HLR. The original HLR sets a level-1 pointer between itself and
the HLR associated with the new MA. The level-1 pointer is also setup
between the new HLR and new MA. Thus, with every movement outside a MA, two
level-1 pointers are set. Thus, k steps outside the MA would result in 2k
level-1 pointers being set, which would cost
\begin{figure}[htb]
\epsfig{file=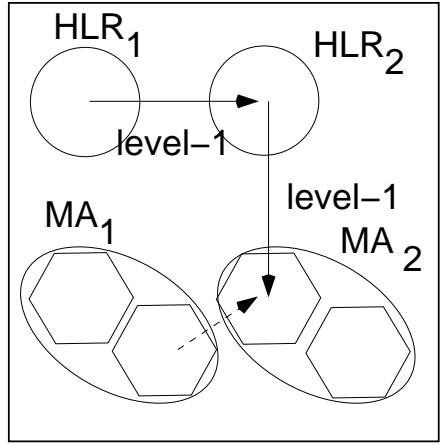, width=2.5in}
\vskip -0.5cm
\caption{Two level-1 pointers are created when user moves between MAs.
Dashed arrow represents level-2 pointer.}
\end{figure}
\beq
=2kS_1
\eeq
where ${\rm S_1}$ is the cost involved in setting a level-1 pointer.

\subsubsection{Net cost due to movement}
The net cost incurred on taking 'i' steps with probability of remaining in
the same MA being ${\rm P_2}$ and that of crossing being ${\rm P_1}$ can be 
written as
\beq
&=& (aj^2+bj-1)S_2+2kS_1\nonumber\\
&=& (ai^2P_2^2+biP_2-1)S_2+2iP_1S_1\nonumber
\eeq
The expected (average) cost would be 
\beq
M'=\sum_0^\infty \left[(ai^2P_2^2+biP_2-1)S_2+2iP_1S_1\right]\alpha (i)
\eeq
where ${\rm \alpha (i)}$ is the probability that 'i' steps would be taken. 
Simplifying
\beq
M' &=& aP_2^2S_2\sum_0^\infty i^2\alpha (i)
+(bP_2S_2+2P_1S_1)\sum_0^\infty i\alpha (i)-
S_2\sum_0^\infty \alpha (i)\nonumber\\
&=& aP_2^2S_2\int_0^\infty i^2\alpha (i)di
+(bP_2S_2+2P_1S_1)\int_0^\infty i\alpha (i)di-
S_2\int_0^\infty \alpha (i)di\label{econd1}
\eeq
To obtain an parametric expression for M', we would have to compute the
integrals. The definition of ${\rm \alpha (i)}$ demands
\beq
\int_0^\infty \alpha (i)di  = 1\label{cond1}
\eeq
\beq
\int_0^\infty i\alpha (i)di = i_{average} ={\mu \over \lambda} ={1 \over
p}\label{cond2}
\eeq
where 'p' represents the {\it Call to Mobility Ratio, CMR}.
\begin{figure}[htb]
\epsfig{file=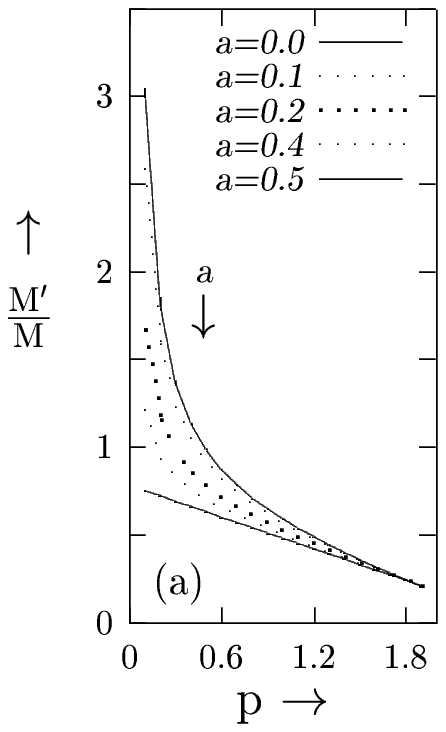, width=1.75in}
\hfil
\epsfig{file=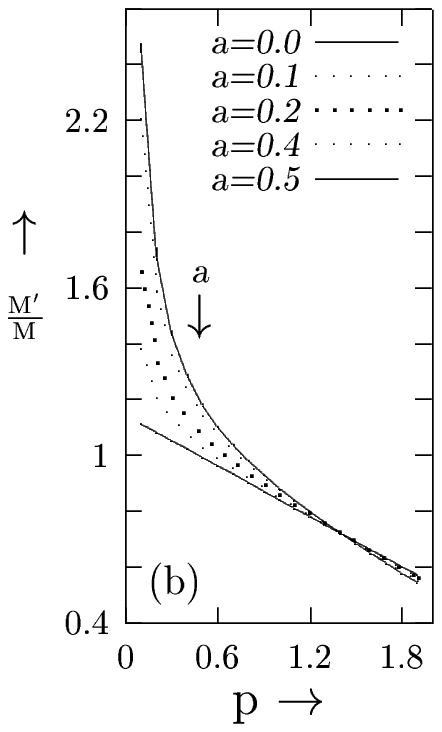, width=1.75in}
\hfil
\epsfig{file=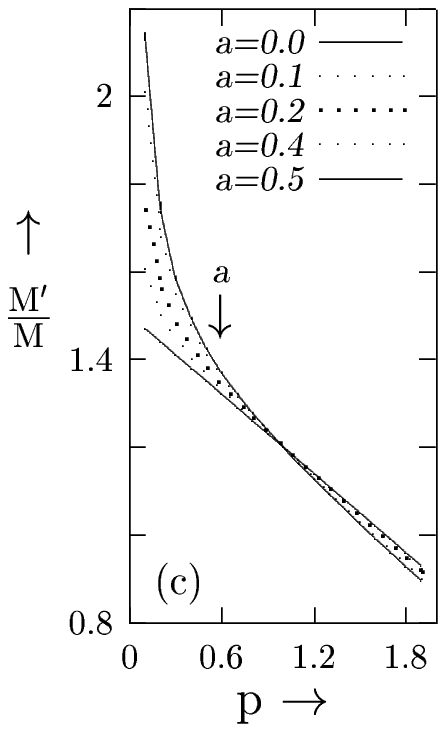, width=1.75in}
\caption{\sl Variation in relative MOVE cost with decreasing probability (a)
${\rm P_2}$= 0.9, (b) ${\rm P_2}$= 0.7 and (c) ${\rm P_2}$= 0.50. The family
of curves is shown for varying coefficient 'a' of eqn(\ref{e6a}).}
\end{figure}
We have
\beq
\int_0^\infty i^2\alpha (i)di = A\int_0^\infty i^2g^idi=
-log(g)\times -{2\over [log(g)]^3}={2 \over p^2}\label{cond5}
\eeq
(detailed derivation is given in the Appendix). On substituting the results of 
eqn(\ref{cond1}), (\ref{cond2}) and (\ref{cond5}), eqn(\ref{econd1}) reduces to
\beq
M' = {2aP_2^2S_2\over p^2}+{(bP_2S_2+2P_1S_1)\over p}-S_2\label{econd2}
\eeq
This is the cost incurred by MOVE (mobility) of the user. Assuming the basic 
MOVE cost to be unity (i.e. x=1), then
\beq
M={x \over p}={1 \over p}\nonumber
\eeq
Hence, the relative MOVE cost would be given as
\beq
{M' \over M}= {2aP_2^2S_2\over p}+(bP_2S_2+2P_1S_1)-pS_2\label{econd3}
\eeq
Since setting up level-1 pointers is always more costlier then setting up
level-2 pointer, ${\rm S_1=\kappa S_2}$ where ${\rm \kappa >1}$. The relative 
cost involved in MOVE can now be written as 
\beq
{M' \over M}=\left[{2aP_2^2\over p}+(bP_2+2\kappa
P_1)-p\right]S_2\label{econd4}
\eeq
\par Figure 5 shows the variation in the expected cost on a MOVE action with
'p', the CMR. The figure shows three cases for different values of ${\rm
P_2}$, the probability of the user moves within the same MA. As expected,
the cost is less for the situation where chances of cross-over is less. This
is also evident from the slope of the curves. Notice that the curves of each
set intersect, however at different values of 'p'. The curves of fig 3 also
intersected at j=1 and the point physically represented the mobile user's
first step. The physical significance of 'j' is distributed to 'p' via ${\rm
P_2}$ in our mathematics. Hence, any interpretation of the curves of fig 5
should be from the point of intersection and towards p=0. Decreasing value
of 'p' represents increasing steps taken by the user or in other words
increasing mobility.

\subsection{Locate}
\begin{figure}[htb]
\epsfig{file=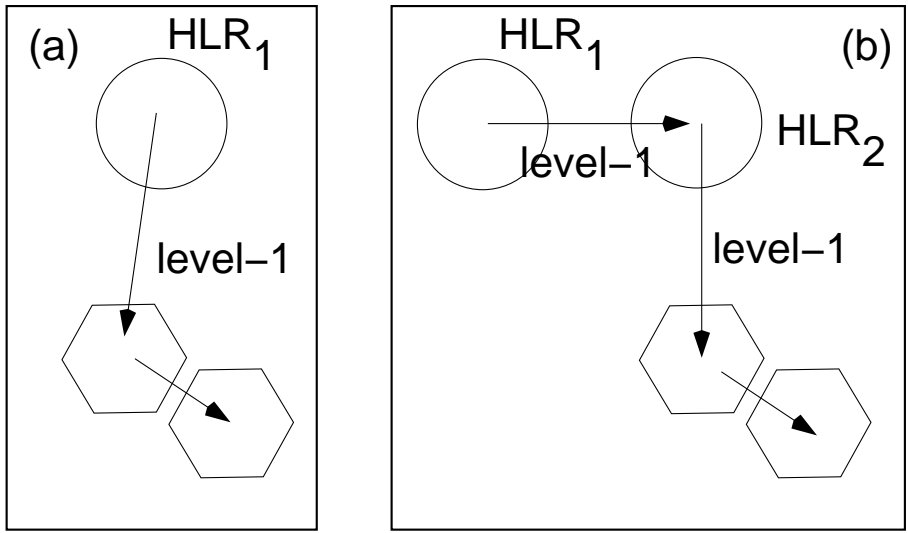, width=5in}
\vskip -0.5cm
\caption{Two possible methods that pointer would be used for locate
(Tracking).}
\end{figure}
The locate activity of the service provider only takes place when a call is
made. Since the level-2 pointer is always being updated to the
shortest length (between VLR occupied at time of last call and the latest VLR
occupied, irrespective of however many steps were taken), there are only two 
possible scenarios for locating a user in this model (see fig 6). The average cost 
incurred in tracing a user would hence be given as
\beq
F'=F+\sum_{i=0}^\infty[iP_2(T_1+T_2)+iP_1(2T_1+T_2)]\alpha(i)
\eeq
where 'F' is the basic cost of a trace (we idealize as =1), ${\rm T_1}$ the
cost of ${\rm level_-1}$ tracing and ${\rm T_2}$ that of ${\rm level_-2}$
tracing. Simplifying, we have
\beq
F' &=& F+[P_2(T_1+T_2)+P_1(2T_1+T_2)]\int_{i=0}^\infty
i\alpha(i)di\nonumber\\
{F' \over F} &=& 1+{[P_2(T_1+T_2)+P_1(2T_1+T_2)] \over p}
\eeq
\begin{figure}[htb]
\epsfig{file=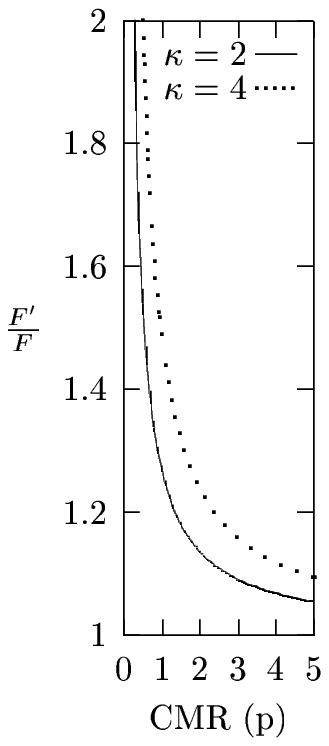, width=1.5in}
\hfil
\epsfig{file=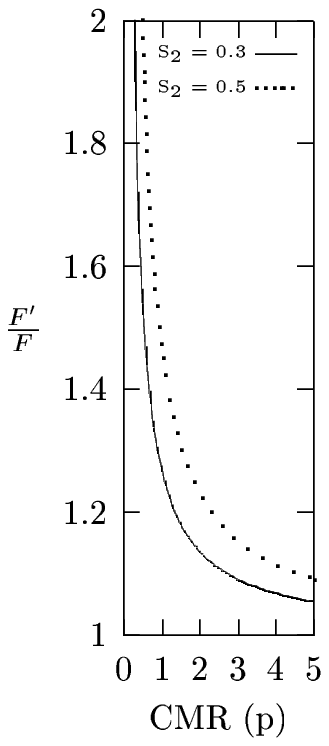, width=1.5in}
\hfil
\epsfig{file=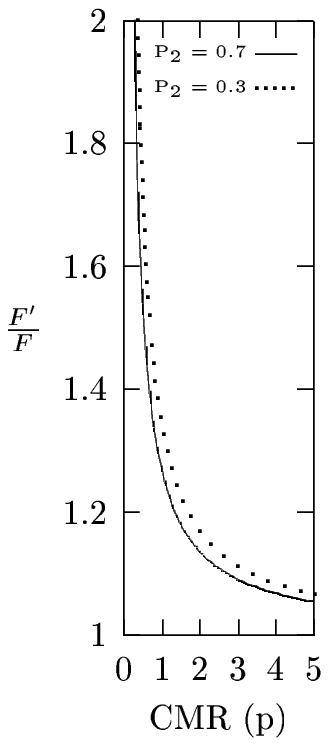, width=1.5in}
\hfil
\epsfig{file=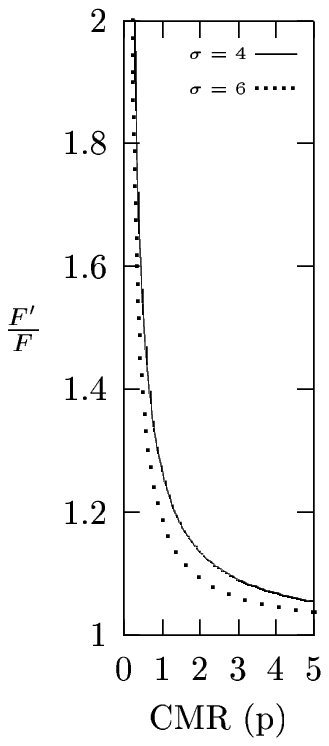, width=1.5in}
\caption{\sl Variation in relative LOCATE cost as a function of ${\rm
\kappa}$, ${\rm S_2}$ , ${\rm P_2}$ and ${\rm \sigma}$.}
\end{figure}
The cost incurred for locating would be less than ${\rm S_1}$ and ${\rm S_2}$
since setting up of pointers involves to and fro  communication between VLRs 
in terms of
registration, de-registration and acknowledgement. Thus, ${\rm S_1=\sigma_1
T_1}$ and ${\rm S_2=\sigma_2 T_2}$. Thus,
\beq
{F' \over F} &=& 1+\left({1\over p}\right)\left[P_2\left({S_1\over \sigma_1}+{S_2\over
\sigma_2}\right)+P_1\left({2S_1\over \sigma_1}+{S_2\over
\sigma_2}\right)\right]\nonumber\\
&=& 1+\left({S_2\over p}\right)\left[P_2\left({\kappa \over \sigma_1}+{1\over
\sigma_2}\right)+P_1\left({2\kappa \over \sigma_1}+{1\over
\sigma_2}\right)\right]\nonumber
\eeq
From the algorithm, it is evident that the number of communications during
pointer set up is same in level-1 and level-2, hence ${\rm
\sigma_1=\sigma_2}$ (where ${\rm \sigma_1, \sigma_2 \gg 1}$). We now write
\beq
{F' \over F} &=& 1+\left({S_2 \over \sigma p}\right)\left[P_2(\kappa
+1)+P_1(2\kappa+1)\right]
\nonumber\\
&=& 1+\left({S_2 \over \sigma p}\right)[1+2\kappa-\kappa P_2]\label{eq20}
\eeq

Eqn(\ref{eq20}) shows that the cost in locating an user's position depends
on the parameters ${\rm S_2}$, ${\rm \kappa}$, ${\rm P_2}$ and ${\rm
\sigma}$. The graphs of fig 7 explicitly exhibits the dependence of relative
locate cost (${\rm F'/F}$) on these parameters. The variations are self
explanatory. However, the important point to be noted from fig 7 is the
strong influence of ${\rm \kappa}$ (in turn ${\rm S_1}$) and ${\rm S_2}$ have 
on ${\rm F'/F}$.

\subsection{Total Cost in Location Tracking and Location Updating}
The total cost of operation would be sum of cost incurred in location
updating and tracking,
\beq
C_F &=& M'+F'\nonumber\\
&=& {2aP_2^2S_2\over p^2}+{(bP_2S_2+2\kappa P_1S_2)\over p}-S_2+
1+\left({S_2 \over \sigma p}\right)[1+2\kappa-\kappa P_2]\label{e29}
\eeq
The cost incurred between two calls in the basic mode of operation is given
as
\beq
C_B &=& M+F={x \over p}+F\nonumber\\
&=&{1 \over p}+1=\left({1+p \over p}\right)\label{e29a}
\eeq
\begin{figure}[htb]
\epsfig{file=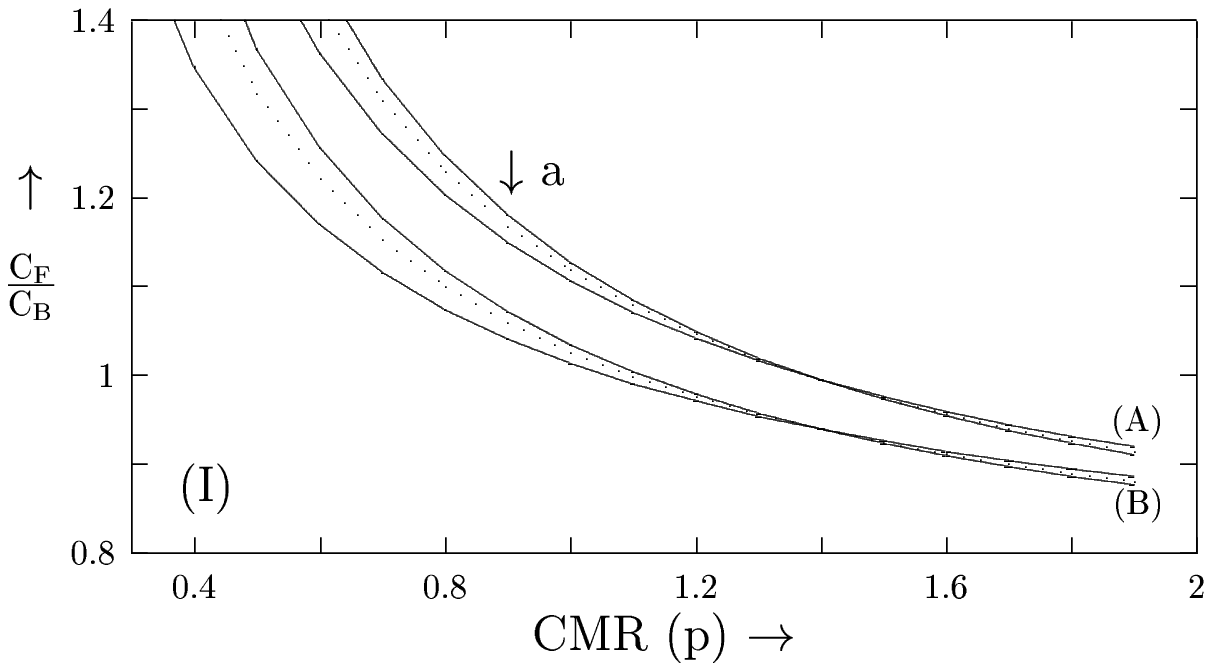, width=3.5in}
\epsfig{file=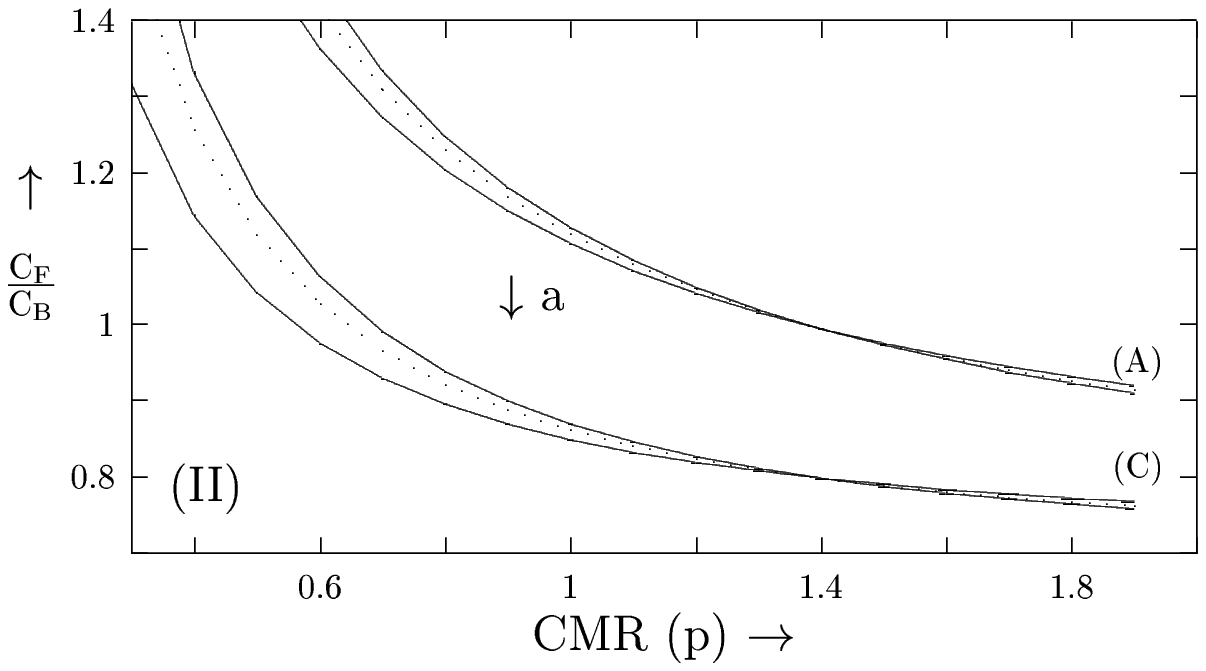, width=3.5in}
\caption{\sl Variation in total cost incurred for both location updating and
location tracking. Curve (A) of both graphs have ${\rm \kappa}$=4 and ${\rm
\sigma}$=5. While curve (B) has ${\rm \sigma}$=10 and cuvre (C)
has ${\rm \kappa}$'s value at 2. Thus, graph I compares the dependence of
cost on ${\rm \sigma}$ and graph II compares the dependence on ${\rm
\kappa}$.}
\end{figure}
For comparing the total cost of the present model w.r.t. the basic model, we
find the ratio ${\rm C_F/C_B}$ using eqn(\ref{e29}) and eqn(\ref{e29a}). We have
\beq
{C_F \over C_B} = 
{2aP_2^2S_2\over p(1+p)}+{(bP_2S_2+2\kappa P_1S_2)\over 1+p}+{p(1-S_2)\over
1+p}+\left[{S_2 \over \sigma (1+p)}\right](1+2\kappa-\kappa P_2)
\eeq
Figure 8 shows the variation of the total cost incurred in the location
management for the proposed TLSCP scheme as compared to the basic operation
mode. Figure 8(I) shows that with increase in ${\rm \sigma}$, the cost of
location management decreases. This is a direct result of decrease in
location tracking cost (${\rm T=S/\sigma}$, see fig 6). However, as seen
from fig 8(II), the cost of location management depends more strongly on 
${\rm \kappa}$'s value, where a lower ${\rm \kappa}$ implies that the cost
of location updating (setting of level-1 pointers) decreases. Also notice
the larger section of curves lieing below (${\rm C_F/C_B}$=) 1 on decreasing
${\rm \kappa}$. This region (${\rm C_F/C_B \leq 1}$) indicates region of
operation where TLSCP scheme works out to be cheaper than the basic
operation scheme. Thus, even for an increased mobility (low p), the cost of
operation in TLSCP scheme can be made low by decreasing the value of ${\rm
\sigma}$ and ${\rm \kappa}$.

\begin{figure}[htb]
\epsfig{file=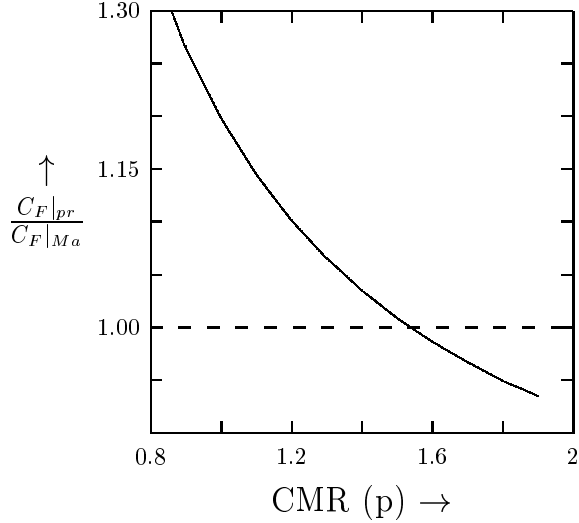, width=3in}
\vskip -0.25cm
\caption{Performance analysis of TLSCP strategy w.r.t. Ma and Fang's
\cite{c15} strategy.}
\end{figure}
\section{Discussion}
The analysis of the TLSCP strategy shows that despite the increase of
localized signal traffic generated in a MA, the cost involved in locating a
user is low. This effectively brings down the cost of operation for users
with low call to mobility ratio. Efficient programming to bring down the
cost in setting up level-1 pointers (making ${\rm \kappa}$ low) would enable
the operator to expand the limit of "the low call to mobility ratio".
Profiling a user and efficiently switching from TLSCP strategy to TLPF (Two
Level Pointer Forwarding \cite{c15}) strategy would bring down the cost of
operation significantly. Figure 9 compares the costing of the two
strategies. This in principle should be easy due to the broad
similarity in the two strategies.

\section{Conclusion}
The present article proposes a Two level single chain pointer forwarding strategy 
for location management. When a mobile user moves from one RA to another, it
is accompanied by a pointer formation between a local anchor (which is an RA
the user occupied when he last received a call) and the new RA in which the
user is located. An increase in cost of MOVE action (various signals
exchanged between RA's and computer information updating) drastically
reduces the cost involved in tracing (Locating) an user to transfer a call.
If the cost incurred in forming a {\it level-2} pointer, i.e. ${\rm S_2}$
can be reduced by minimizing signal transfers required to update computer
information, the total cost for location updating and location tracking
(${\rm C_F}$) would be reduced. This would bring down the operating cost of
the service provider. In the present model since a MA can
manage larger RA's in it, increasing number of RA's under a MA decreases the
probability of user crossing over MA's. This again strongly contributes to
the decrease in operating cost. The present strategy (TLSCP) out performs
the basic strategy. However, it only betters Ma and Fang (TLPFS)\cite{c15} 
strategy for low mobility. The model fails in decreasing cost for a user with 
large mobility. It would be interesting to investigate an algorithm in
future works, that switches
from TLSCP model to TLPFS as an increase in user mobility is detected.

\section*{APPENDIX}
Based on the assumption that the residency time of a user in a RA follows Gamma
Distribution, Ma and Fang\cite{c15} derived an expression for ${\rm \alpha (i)}$,
given as
\beq
\alpha (i)={[g^{i-1}(1-g)^2] \over p}=Ag^i\label{cond2f}
\eeq
'g' is the Laplace transform of the Gamma Distribution function. The
constant 'A' and the nature of 'g' can be evaluated by subjecting it to the
conditions given by eq(\ref{cond1}), i.e.  
\beq
\int_0^\infty \alpha (i)di  = 1\nonumber
\end{eqnarray}
using eqn(\ref{cond2f}) in this relation, we have
\begin{eqnarray}
A\int_0^\infty g^idi =- {A \over [log(g)]} = 1\nonumber
\eeq
This gives
\beq
A &=& -log(g)\label{cond3}
\eeq
(this is subjected to the condition ${\rm g<1}$). From eqn(\ref{cond2}) we get
\beq
\int_0^\infty i\alpha (i)di &=& {1 \over p}\nonumber\\
-log(g)\int_0^\infty ig^idi &=& {1 \over p}\nonumber\\
-{1 \over [log(g)]} &=& {1 \over p}\nonumber
\eeq
This gives the functional form of 'g' as {\it exp(-p)} and in turn 
\beq
\alpha (i) = pe^{-pi}\label{cond4}
\eeq
where p is the CMR.



\begin{thebibliography}{99}















\bibitem{c1} Poole I, {\it From Analogue to 3G}, IEEE Communications Engineering,
June/July, 2003, pg 26-29. 

\bibitem{c2} Varshney U., {\it Location management for wireless networks}, Proc. 
Int. Conference on Telecom Systems, March 1999.

\bibitem{c3} EIT/TIA , {\it Cellular Radio Telecommunications Inter sytem 
Operations}, Technical Report IS-41 (Revision B), EIA/TIA, 1991.

\bibitem{c4} ETSI, Digital Cellular Telecommunications System (Phase 2+): 
Mobile Applications Part (MAP) Specification (GSM 09.02 version 7.51 Release ), 
1998.

\bibitem{c5}  Huang T., {\it Overview of GSM: Philosophy and Results}, 
International Journal of Wireless Information Networks, 1994.

\bibitem{c6} Mukerjee A, Saha D and Jha S, {\it Location management in mobile 
wireless networks}, Wieless internet book: technologies, standards and 
application, ISBN:0-8493-1502-6, 2003, p.351-380.

\bibitem{c7}  Chen I. R., Chen T. M. and Lee C., {\it Agent-based 
Forwarding Strategies for reducing location management cost in mobile
networks}, ACM/Baltzer J. Mobile Networks Applications, {\bf 6} (2001), 
pg 105-116.

\bibitem{c8}  Sen S. K., Bhattacharya A. and Das S. K., {\it A selective 
location update strategy for PCS users}, ACM/Baltzer Wireless Networks, {\bf
5} (1999),313-326.

\bibitem{c9}  Wong V. W. S. and Leung V. C. M., {\it Location management for 
next-generation personal communications networks}, IEEE Network, {\bf 14}
(2000), 18-24.

\bibitem{c10}  Sue. K. and Tseng Chieu-Chao, {\it One step pointer forwarding 
strategy for location tracking in distributed HLR environment}, IEEE ACM 
Transactions on Networking, {\bf 6} (1998) 

\bibitem{c11} Jain R. and Lin Y. B., {\it An auxiliary user location strategy 
employing forwarding pointers to reduce network impacts of PCS}, Wireless 
Network, {\bf 1} (1995), 197-210.

\bibitem{c12}  Ho J. and Akylidiz F, {\it Local anchor scheme for reducing 
signaling costs in Personal Communication Networks}, IEEE/ACM Transactions 
Networking, {\bf 4} (1996).

\bibitem{c13} Lin H, Lee S and Lin W, {\it Dyanamic location tracking strategy 
for hot mobile subscribers in personal communications}, Computer Communications, 
{\bf 26} (2003), 1353-1364.

\bibitem{c14} Akylidiz I. F. and Wang W, {\it A dyanamic location management 
scheme for next generation multitier PCS systems}, IEEE Transactions on 
Wireless Communications, {\bf 1} (2002).

\bibitem{c15} Ma W and Fang Y, {\it Two level pointer forwarding strategy for 
location management in PCS Networks}, IEEE Transactions On Mobile Computing, 
{\bf 1} (2002), 32-44.
 
\bibitem{c16} Akylidiz I. F., Ho J. S. M. and Lin Y. B, {\it Movement based 
location update and selective paging for PCS network}, IEEE/ACM Trans. Networking, 
{\bf 4} (1996), 629-638.


\bibitem{c17} Fang Y and Ma W, {\it Mobility management for wireless networks: 
modeling and analysis}, Wireless Communications Systems and Networks, 
ISBN:0-306-48190-1, 2004, pg 473-512.                                              
      
\bibitem{c18} Chen I, Gu B, {\it A comparitive cost analysis of degradable 
location management algorithms in wireless networks}, The Computer Journal, 
British Computer Society, {\bf 45} (2002), 304-319.

\bibitem{c19} Chhaya R. Kant, {\it Two level single chain pointer forwarding 
strategy for location management in PCS networks}, Dissertation submitted for 
fulfillment of M. S. (Software Systems), Birla Institute of Technology and 
Science, Dec (2003). 

\bibitem{c20} Prakash N and Mahapatra A. K., {\it Performance Analysis of two 
level single chain pointer forwarding strategy for location management}, in 
Proceedings of IEEE International Conference on Personal Wireless Conference 
(ICPWC-2005555), Jan 2005, Delhi, pg 394-398.

\bibitem{c21} Tanenbaum A. S., {\it Computer Networks}, Prentice Hall India, 
3rd ed. (1997).

\end{thebibliography}
\end{document}